\documentstyle[epsf]{article}

\begin{document}

\title{Quasinormal modes of \\ maximally charged black holes\thanks{
TIT/HEP-311/COSMO-61}}
\author{Hisashi Onozawa\thanks{E-mail:onozawa@phys.titech.ac.jp},
Takashi Mishima$ ^\ddagger$, Takashi Okamura, \\
and \\
Hideki Ishihara \\
\\
Department of Physics, \\
Tokyo Institute of Technology,\\
Oh-okayama, Meguro, Tokyo 152, Japan\\
\\
$ ^\ddagger$Laboratories of Physics, College of Science and Technology, \\
Nihon University, Narashinodai, Funabashi, Chiba 274, Japan}
\date{December 31, 1995}

\maketitle

\begin{abstract}
A new algorithm for computing the accurate values of 
quasinormal frequencies of extremal Reissner-Nordstr\"{o}m black
holes is presented.
The numerically computed values are 
consistent with the values earlier obtained by Leaver
and those obtained through the WKB method.
Our results are more precise than other results known to date.
We also find a curious fact that the resonant frequencies of
gravitational waves with multi-pole index $l$ coincide with
those of electromagnetic waves with multi-pole index $l-1$ in
the extremal limit.
\end{abstract}

\section{Introduction}

The determination of the quasinormal frequencies of black holes
has for several years been of interest and is
related to experiments which aim at 
detecting gravitational
waves from supernovae or coalescences of binary neutron stars,
which are thought to eventually form a black hole.
In the late stage of the process of black hole formation,
a certain mode of the gravitational wave dominates the
emission. This is called the quasinormal mode of a black hole.

The equations governing oscillations of black holes are derived by
Regge and Wheeler\cite{REG57}, and Zerilli\cite{ZER71}
for the non-rotating uncharged case, Zerilli\cite{ZER74}
and Moncrief\cite{MON} for the non-rotating
charged case and Teukolsky\cite{TEU72} for the rotating uncharged case.
The resonant frequencies of the Schwarzschild black holes 
were first computed
by Chandrasekhar and Detweiler\cite{CHA75a}, who treated it as
a boundary value problem of the 
second order ordinary differential equation of Regge-Wheeler.
Integration of the Regge-Wheeler equation for the quasinormal boundary
condition
is numerically unstable.
To avoid the instability, Leaver\cite{LEA85} presents the 
numerically stable continued fraction
method for Schwarzschild and Kerr black holes. Here,
a quasinormal mode function
corresponds to a minimal solution of the recurrence relations
which are satisfied by the coefficients of a series expansion
of a wave function.
The minimal solution of the three-term recurrence relation
is obtained by the corresponding continued fraction\cite{GAU67}.
The continued fraction method can give the values
of frequencies with high numerical precision
because it uses
no approximation, as is the case in the semi-analytic WKB method developed
in other works\cite{FER84,BOL84,SCH85,IYE87}.
Leaver\cite{LEA90} also generalizes the continued fraction method 
to calculate accurate values of the
quasinormal frequencies of Reissner-Nordstr\"{o}m black holes,
though Kokkotas and Schutz\cite{KOK88} raised the
question of the applicability of the continued fraction method 
to charged black holes.

We are indeed motivated by his paper\cite{LEA90}.
There is another interesting question, that is, whether a continued
fraction method can be used for the maximally charged black holes
not discussed in his paper.
In the extremal limit, the wave equation
has an irregular singular point at
the horizon, which makes the series expansion
of a solution there invalid, and thus
a continued fraction method does not seem to work.
In this paper we show that 
a continued fraction method is applicable 
even to the extremal black holes
if we expand a solution about a suitable ordinary point.
In section II, we derive an eigenvalue equation with 
continued fractions
after expanding a wave function about an ordinary point
to determine the quasinormal frequencies of the extremal
Reissner-Nordstr\"{o}m black hole.
In section III, we summarize the numerical results and
find an interesting coincidence between the resonant frequencies
of gravitational waves with multi-pole index $l$ and
those of electromagnetic waves with multi-pole index $l-1$
in the limit of maximal charge.
Section IV is devoted to conclusions and discussion.

\section{Eigenvalue Equation for Quasinormal Modes}

Leaver obtained the recurrence relations of Schwarzschild, Kerr\cite{LEA85},
and Reissner-Nordstr\"{o}m black holes\cite{LEA90} 
by expanding solutions about the
event horizon and then solved the characteristic equations with
a continued fraction to get these resonant frequencies accurately.
However his method is not valid when a black hole is
extremal. The wave equation has an irregular singular point at the horizon
about which 
the expansion of the solution is not available
because 
a radius of convergence for a series expansion about an irregular
singular point generally vanishes.
Therefore we have to expand the solution about a suitable ordinary
point in the extremal black holes.

We start from the Zerilli-Moncrief equation for perturbations
on Reissner-Nordstr\"{o}m backgrounds of mass $M$ and charge $Q$
written as
\begin{equation}
  \left[ \frac{d^2}{dr^2_*}+ \omega^2 -V_s(r) \right] Z_s(r) =0,
  \label{eq:rw}
\end{equation}
where
\begin{eqnarray}
  \frac{dr}{dr_*} & = & \frac{\Delta}{r^2}, \\
  V_s    & = & \frac{\Delta}{r^5} \left[ A r 
                  -q_s +\frac{p_s}{r}\right], \\
  \Delta & = & r^2-2Mr+Q^2,\\
  q_0    & = &  2, \\
  q_1    & = & 3 M- \sqrt{9 M^2+4Q^2 (A-2)},\\
  q_2    & = & 3 M+ \sqrt{9 M^2+4Q^2 (A-2)},\\
  p_0    & = & 2, \\
  p_1    & = &  p_2 = 4,\\
  A      & = & l(l+1),
\end{eqnarray}
and $s=0,1,2$ 
are for a massless scalar field, an odd parity electromagnetic perturbation,
and an odd parity gravitational perturbation, respectively.
The new parameter $p_s$ is introduced here in order to deal with the
case of the scalar field as well.
Note that the definitions of $q_1$ and $q_2$ are
exchanged with each other from the
book of Chandrasekhar\cite{CHA83}, though the most of other
notations follow this book.
The scaling of $t$ and $r$ in this paper is such that $c=G=M=1$.
We also use $\rho=-i\omega$ as a frequency variable for convenience.

Now we follow Leaver's procedure of getting a recurrence
relation which plays a central role in a continued fraction method.
The tortoise coordinate in the extremal limit becomes
\begin{equation}
  r_* = r+ \log (r-1)^2 - \frac{1}{r-1},
\end{equation}
which makes a difference in the basic equations
between extremal and non-extremal black holes.
The Zerilli-Moncrief equation (\ref{eq:rw}) is
therefore rewritten in terms of $r$-coordinate as follows,
\begin{eqnarray}
&&  r (r-1) \frac{d^2 Z_s}{dr^2} + 2 \frac{dZ_s}{dr}
        \nonumber\\
&& - \left[ \frac{\rho^2 r^5}{(r-1)^3}+\frac{Ar}{r-1}
        - \frac{q_s}{r-1}+\frac{p_s}{r(r-1)}\right]Z_s =0.
  \label{RW}
\end{eqnarray}
This equation has irregular singular points
at $r=1$ and $r=\infty$ as we mentioned in the previous section, which
makes the series expansion about the horizon invalid because
it is hopeless that the series expansion is convergent.
Hence it is necessary to improve Leaver's 
treatment of obtaining the quasinormal modes of the extremal black holes.

The boundary conditions of a quasinormal wave function  are given by
\begin{eqnarray}
  Z_s \sim e^{-\rho r_*} & \mbox{as} & 
r_* \rightarrow \infty, \\
  Z_s \sim e^{\rho r_*}  & \mbox{as} & r_* \rightarrow  -\infty , 
\end{eqnarray}
which mean there only exists purely outgoing wave at the infinity and 
purely ingoing wave at the black hole horizon.
These are also transformed to 
\begin{eqnarray}
Z_s \sim r^{-2 \rho}e^{-\rho r}  & \mbox{as} & r \rightarrow \infty, \\
Z_s \sim e^{\rho r} (r-1)^{2\rho} e^{-\rho \frac{1}{r-1}} 
& \mbox{as} & r \rightarrow 1,
\end{eqnarray}
which determine a pre-factor of a quasinormal wave function
in the following form
\begin{equation}
  Z_s=e^{\rho r} e^{-\rho \frac{1}{r-1}} (r-1)^{2\rho} r^{-4 \rho}
     e^{-2\rho (r-1)} \sum_{n=0}^{\infty}  
    a_n u^n,
 \label{eq:expansion}
\end{equation}
where $u$ is specified by a point where we expand a solution.
We use a series expansion about an ordinary point,
which has a finite radius of convergence at least up to the nearest
singular point. Thus, we use $u =(r-2)/r$ instead of 
$u =(r-1)/r$ which is used in Ref.\cite{LEA90}.
In our choice of $u$ 
the black hole horizon and infinity correspond to $u=-1,1$, respectively.
The distances from the point $u=0$ to two singular points $u=-1,1$
are equal, which can make
us possible to examine the boundary conditions at both sides simultaneously.
It is clear that the quasinormal mode boundary conditions
are satisfied if and 
only if $\Sigma a_n u^n$ converges at both boundaries, horizon and
infinity.
This is equivalent to the condition that 
both $\Sigma a_{n}$ and $\Sigma (-)^n a_{n}$ are
finite, that is, both $\Sigma a_{2n}$ and $\Sigma a_{2n+1}$ are convergent
simultaneously.
To examine the convergence of each summation,
we first obtain the recurrence relation of coefficients $a_n$
of the expansion (\ref{eq:expansion}),
and then attempt to obtain the recurrence relation
only for $b_n =a_{2n+1}$ and that of $c_n =a_{2n}$.
The quasinormal mode boundary conditions
are satisfied when the odd and even sequence are minimal
solutions of the corresponding recurrence relations, respectively.

After substituting the above series expansion (\ref{eq:expansion})
into Eq.(\ref{RW})
we obtain a five-term recurrence relation in the following,
\begin{eqnarray}
\alpha_1 a_2 + \gamma_1 a_0 &=& 0, \label{eq:a20} \\
\alpha_2 a_3 + \gamma_2 a_1 + \delta_2 a_0 &=& 0, \label{eq:a31}\\
\alpha_n a_{n+1} + \beta_n a_{n} + \gamma_n a_{n-1} + \delta_n a_{n-2} +\epsilon_n a_{n-3} &=& 0, 
(n \ge 3), \label{eq:rec}
\end{eqnarray}
where the recurrence coefficients are given by
\begin{eqnarray}
\alpha_n &=& n^2+n, \\
\beta_n &=& 0, \\
\gamma_n &=& -4A-2n^2-p_s+2q_s+n(2-24\rho)+12\rho-64\rho^2,\\
\delta_n &=& 2(p_s-q_s),\\
\epsilon_n &=& n^2-p_s-12\rho+16\rho^2+n(-3+8\rho).
\end{eqnarray}

For the scalar field, we can find that 
$\beta_n=\delta_n=0$ in the above relation,
which means an odd sequence and even sequence are completely decoupled with
each other, because
the Zerilli-Moncrief equation in terms of $u$ 
is symmetric under the transformation $u \rightarrow -u$.
Therefore it has only symmetric or anti-symmetric
solutions as quasinormal mode functions.
This is the same as the case of a quantum mechanical system with 
a reflection symmetry,
where a wave function of energy eigenstate must be either symmetric 
or anti-symmetric.
This makes the calculation of quasinormal frequencies of 
the scalar field easy.
Each sequence is characterized by
the three-term recurrence relation (\ref{eq:rec}) starting from the
two-term relation (\ref{eq:a20}) or (\ref{eq:a31}).
We can obtain two independent characteristic equations below similar
to one for the quasinormal frequencies of Schwarzschild black holes,
\begin{eqnarray}
 \frac{a_2}{a_0} = - \frac{\gamma_1}{\alpha_1} = 
  \frac{-\epsilon_2}{\gamma_2-}\frac{\alpha_2\epsilon_4}
  {\gamma_4-}\frac{\alpha_4\epsilon_6}{\gamma_6-} \cdots, 
    \label{eq:eigen-even} \\
 \frac{a_3}{a_1} = - \frac{\gamma_2}{\alpha_2} = 
  \frac{-\epsilon_3}{\gamma_3-}\frac{\alpha_3\epsilon_5}
  {\gamma_5-}\frac{\alpha_5\epsilon_7}{\gamma_7-} \cdots.
  \label{eq:eigen-odd}
\end{eqnarray}
These two equations are for the modes 
of even-numbered and those of odd-numbered overtone, respectively.

For gravitational or electromagnetic quasinormal modes,
we have to develop a little more complicated way
because the presence of $\delta_n$-term
makes odd and even sequences dependent on each other,
though it seems that they are not much correlated 
for large $n$ due to the fact that
the coefficient $\delta_n$ becomes relatively small
as $n$ increases compared with other terms.
We first need to eliminate the odd or even sequence to examine
the convergence of the summation of each sequence.
This procedure is presented in Appendix A.
For $b_n$ a five-term recurrence relation below is obtained,
\begin{eqnarray}
\hat{\alpha}_1 b_2 + \hat{\beta}_1 b_1 + \hat{\gamma}_1 b_0 &=& 0, \\
\hat{\alpha}_2 b_3 + \hat{\beta}_2 b_2 + \hat{\gamma}_2 b_1 + \hat{\delta}_2 b_0 &=& 0, \\
\hat{\alpha}_n b_{n+1} + \hat{\beta}_n b_n + \hat{\gamma}_n b_{n-1} + \hat{\delta}_n b_{n-2} + \hat{\epsilon}_{n} b_{n-3}&=& 0, 
(n \ge 3) \label{eq:recodd},
\end{eqnarray}
and for $c_n$ we get a following relation,
\begin{eqnarray}
\bar{\alpha}_2 c_3 + \bar{\beta}_2 c_2 + \bar{\gamma}_2 c_1 + \bar{\delta}_2 c_0 &=& 0, \\
\bar{\alpha}_n c_{n+1} + \bar{\beta}_n c_n + \bar{\gamma}_n c_{n-1} + \bar{\delta}_n c_{n-2} +\bar{\epsilon}_{n} c_{n-3}&=& 0, 
(n \ge 3) \label{eq:receven},
\end{eqnarray}
where each recurrence coefficient is also given in Appendix A.

As mentioned previously, the quasinormal mode boundary conditions are
satisfied if and only if both sequences are minimal solutions of 
recurrence relations simultaneously.
In order to use a continued fraction method,
both five-term recurrence relations are transformed to
three-term recurrence relations
by successive Gaussian elimination steps
which are exhibited in Appendix B.
Then, the ratios of the first two terms of both
sequences for minimal solutions are determined by
the corresponding continued fractions in the following,
\begin{eqnarray}
 \frac{a_4}{a_2} = \frac{c_2}{c_1} = 
  \frac{-\bar{\gamma}''_2}{\bar{\beta}''_2-}\frac{\bar{\alpha}''_2\bar{\gamma}''_3}
  {\bar{\beta}''_3-}\frac{\bar{\alpha}''_3\bar{\gamma}''_4}{\bar{\beta}''_4-} \cdots , \label{eq:cf42}\\
 \frac{a_3}{a_1} = \frac{b_1}{b_0} = 
  \frac{-\hat{\gamma}''_1}{\hat{\beta}''_1-}\frac{\hat{\alpha}''_1\hat{\gamma}''_2}
  {\hat{\beta}''_2-}\frac{\hat{\alpha}''_2\hat{\gamma}''_3}{\hat{\beta}''_3-} \cdots ,\label{eq:cf31}
\end{eqnarray}
where the quantities with double-prime are recurrence coefficients of
transformed three-term recurrence relations in Appendix B.

In contrast to scalar field case, the even and odd sequences are
coupled each other through the non-vanishing $\delta_n$ terms.
Dividing Eq.(\ref{eq:rec}) for $n=3$ by Eq.(\ref{eq:rec}) for $n=5$,
we get the relation between the ratio of the first two terms of
odd sequence and that of even sequence as follows,
\begin{eqnarray}
  \frac{a_3}{a_1} & = &{
  \alpha_5 \frac{\displaystyle{a_6}}{\displaystyle{a_4}}+\gamma_5+\epsilon_5 \frac{\displaystyle{a_2}}{\displaystyle{a_4}}
 \over  \alpha_2 \frac{\displaystyle{a_4}}{\displaystyle{a_2}}+\gamma_3+\epsilon_3 \frac{\displaystyle{a_0}}{\displaystyle{a_2}} }
   \frac{a_4}{a_2} \label{eq:qnm_e} \nonumber \\
& = &{  - \frac{\displaystyle{\alpha}_5}{\bar{\displaystyle{\alpha}}''_2}
   (\bar{\beta}_2'' + \frac{\displaystyle{a}_2}{\displaystyle{a}_4}
\bar{\gamma}''_2)
+\gamma_5+\epsilon_5 \frac{\displaystyle{a_2}}{\displaystyle{a_4}}
 \over  \alpha_3 \frac{\displaystyle{a_4}}{\displaystyle{a_2}}+\gamma_3-\epsilon_3 \frac{\displaystyle{\alpha_1}}{\displaystyle{\gamma_1}} }
   \frac{a_4}{a_2}, \label{eq:qnm_ern}
\end{eqnarray}
where the second equality comes from 
Eq.(\ref{eq:a20}) and the transformed three-term relation
(\ref{eq:rec-e}) for $n=2$.
The quasinormal mode boundary conditions are satisfied when
the continued fractions (\ref{eq:cf42}) and (\ref{eq:cf31})
satisfy Eq.(\ref{eq:qnm_ern}).
This is the characteristic equation for the quasinormal frequencies of
the extremal Reissner-Nordstr\"{o}m black hole.

Eq.(\ref{eq:qnm_e}) can be shifted an arbitrary number of
times, $n$, to yield an equation for the ratio of successive terms
of higher $n$:
\begin{equation}
  \frac{a_{n+3}}{a_{n+1}}={
  \alpha_{n+5} \frac{\displaystyle{a_{n+6}}}{\displaystyle{a_{n+4}}}+\gamma_{n+5}+\epsilon_{n+5} \frac{\displaystyle{a_{n+2}}}{\displaystyle{a_{n+4}}}
 \over  \alpha_{n+2} \frac{\displaystyle{a_{n+4}}}{\displaystyle{a_{n+2}}}+\gamma_{n+3}+\epsilon_{n+3} \frac{\displaystyle{a_{n}}}{\displaystyle{a_{n+2}}} }  \frac{a_{n+4}}{a_{n+2}}. \label{eq:qnm_e_n}
\end{equation}
Eq.(\ref{eq:qnm_e}) and Eq.(\ref{eq:qnm_e_n}) are 
completely equivalent 
since every solution to (\ref{eq:qnm_e}) is also a solution 
to (\ref{eq:qnm_e_n})
when we substitute a continued fraction into each ratio of $a_i/a_{i-2}$,
and vice versa.
Eq.(\ref{eq:qnm_e_n}) can be used 
as a check of results obtained from Eq.(\ref{eq:qnm_e}).

\section{Numerical Results}

The eigenvalue equations 
(\ref{eq:eigen-even}), (\ref{eq:eigen-odd}) and (\ref{eq:qnm_e})
derived in the previous chapter are
solved using the HYBRD subroutine distributed in MINPACK libraries.
The three least-damped modes of 
$(s,l)=(2,2)$, $(2,3)$, $(2,4)$, $(1,1)$, $(1,2)$,
$(1,3)$, $(0,0)$, $(0,1)$, $(0,2)$
for the extremal Reissner-Nordstr\"{o}m black hole are 
listed in Table 1.
We compare our results with Leaver's results\cite{LEA90}
of nearly maximally charged case (99.99\% charged),
which should be divided by two because of the difference of
our scaling from his.
These modes appear as complex conjugate pairs in $\rho$
because the eigenvalue equations are characterized by only real coefficients.
We also compare them with the third order WKB quasinormal
frequencies of the extremal Reissner-Nordstr\"{o}m black hole.
The WKB quasinormal frequencies
are obtained using the same formula of Kokkotas and Schutz\cite{KOK88},
who calculated the frequencies of charged black holes
but did not show those of maximally charged black holes in the paper.

Our results are in agreement with 
other results obtained by Leaver and those obtained through the WKB method
within the accuracy of a few percent.
The difference between WKB frequencies and ours is ascribed
to the fact that the WKB method gives only the approximate
values.
Indeed, the tendency that the discrepancy grows with the mode number
suggests the breakdown of the WKB approximation for higher $n$.
The difference between Leaver's results and ours is
from the numerical error caused by the breakdown of his series expansion
at the limit of maximal charge.
Thus we are sure that our quasinormal frequencies of
the extremal Reissner-Nordstr\"{o}m black hole are the most accurate.
Our results and his are plotted in Fig.1.

Notice that the quasinormal frequencies of $(s,l)=(2,2)$ and
those of $(1,1)$ approach each other with increasing $Q$ and coincide
in the extremal limit.
We show this curious coincidence in the limit of
maximal charge explicitly in the Fig.2.
The trajectories of the first four lowest modes of
the third order WKB gravitational
quasinormal frequencies of $l=2,3,4,5$ are plotted.
These quasinormal frequencies coincide with the frequencies of 
electromagnetic wave of $l=1,2,3,4$ in the
limit of maximal charge, respectively.
The first order WKB quasinormal frequencies are given by
$\omega_n^2 = {V_{peak}} - i (n+1/2) \sqrt {-2V_{peak}''}$
using the Zerilli-Moncrief Potential $V$,
where a subscript $n$ is a mode number.
The differences of $V_{peak}$ and $V_{peak}''$ between $V_1$ for $l=1$
and $V_2$ for $l=2$
are calculated to
$\Delta V_{peak} /V_{peak} \simeq 2.79565 \times 10^{-16}$ and  
$\Delta V_{peak}''/V_{peak}'' \simeq  5.46819 \times10^{-15}$, respectively.
This explains the coincidence of a few least-damped quasinormal
frequencies of these two modes.
Even for higher overtone modes $n \sim 30$, however,
we numerically find a remarkable
coincidence in the quasinormal frequencies
of gravitational waves of $l$
and those of electromagnetic waves of $l-1$ 
for a wide range of multi-pole indices, $2 \le l \le 10$,
through our method.

These two potentials, $V_1$ with multi-pole index $l$ and
$V_2$ with multi-pole index $l+1$, indeed give the same transmission
and reflection amplitude for each $\sigma$, whose
evidence is presented in Appendix C.
It implies that the quasinormal frequencies for both potentials
are identical.

\section{Conclusions and Discussion}

We improve the continued fraction method of computing quasinormal frequencies
of charged black holes in order to determine the quasinormal
frequencies in the extremal limit
and solve our formulae of 
eigenvalue equations by using a root search program.
Our results of quasinormal frequencies for the maximally charged black holes
are consistent with any other previous results\cite{LEA90,KOK88}
and are the most accurate of all.

Our procedure is summarized as follows.
We expand the solution about the ordinary point and get the 
recurrence relation of the expansion coefficients.
In our method we have to examine the boundary conditions
at both boundaries.
Then, we need to divide the sequence into
an odd sequence and an even sequence
to examine convergence of both series.
The convergence of the series at both sides occurs 
if and only if each sequence is minimal.
The ratio of successive terms of each sequence
is given by a continued fraction.
Consequently,
the eigenvalue equation is constructed by substituting
the continued fraction into the ratio of successive recurrence relations.

The applicability of our method is free from types of singularities
of the equation, in contrast to Leaver's method which is not
useful when the equation has two irregular singular points at both
boundaries.
Though we can use our method to
determine quasinormal frequencies of other black holes such as the
Schwarzschild, non-extremal Reissner-Nordstr\"{o}m and Kerr black holes,
we actually did not follow this procedure in these black holes.
For these black holes,
his method is easier to obtain the quasinormal frequencies
because the recurrence relation and the eigenvalue equation 
are simpler than ours.
In our method, it is generally difficult to give the explicit forms of 
the recurrence relations of both even and odd sequences.
In that case we will have to eliminate the
even or odd sequence by numerical elimination steps.
For the extremal Reissner-Nordstr\"{o}m black hole,
the recurrence relation, Eq.(\ref{eq:rec}),
is not so complicated in that it has no $\beta_n$-term
and furthermore
$\delta_n$-term has no dependence on $n$, which makes our procedure
rather easy to accomplish.

We believe our method can be also used for the extremal Kerr black hole
which has an irregular singular point at the horizon in the
equation for radial part.
Some authors\cite{TEU74,DET80,SAS90}
obtain the series of infinite numbers of quasinormal
frequencies of the extremal Kerr black hole,
which accumulate onto the critical frequency.
Leaver also finds a tendency of an accumulation 
of quasinormal frequencies
when the Kerr parameter approaches the maximum value.
It is interesting to investigate what will happen to the
distribution of quasinormal frequencies in the limit of maximal rotation,
where Leaver's method is not valid.
We are in preparation for this problem through our method
of computing quasinormal frequencies.

We accidentally find a curious coincidence in the quasinormal
frequencies of gravitational perturbations with multi-pole index $l$
and those of electromagnetic perturbations with $l-1$.
These two modes are completely decoupled but they have the same amplitudes
of transmission and reflection.
In that case only the difference is a phase shift of transmitted or
reflected wave, that means the quasinormal frequencies for both modes are
identical.
The situation is very similar to the coincidence
of the transmission and reflection amplitudes of
odd parity perturbations with those of even parity perturbation\cite{CHA83}.
At present there is no easy way to understand the hitherto
unobserved coincidence but it is interesting that 
it occurs only in the extremal limit,
where the black hole may have an unknown symmetry.

\section*{Acknowledgments}

We would like to thank to Prof. A.~Hosoya for his continuous encouragement.
The research is supported in part by the Scientific Research Fund of the
Ministry of Education.

\section*{Appendix A}

Here we summarize a procedure to obtain a recurrence relation
for the odd sequence and that for the even sequence of $a_n$.
First we write down the recurrence relation, Eqs.(\ref{eq:a20}),
(\ref{eq:a31}) and ({\ref{eq:rec}),
in the matrix expression
and separate the odd sequence from the even sequence as follows,
\begin{eqnarray}
\left( 
\begin{array}{lllll}
  \gamma_1 & \alpha_1 & & &  \\
  \epsilon_3  & \gamma_3 & \alpha_3 & &  \\
             & \epsilon_5 & \gamma_5 & \alpha_5  & \\
    & & \ddots & \ddots &  \ddots
\end{array}
\right)
\left(
\begin{array}{l}
a_0 \\
a_2 \\
a_4 \\
\vdots
\end{array}
\right)
= -
\left(
\begin{array}{lllll}
0 & & & & \\
  & \delta & & & \\
  & & \delta & & \\
  & & & \ddots &
\end{array}
\right)
\left(
\begin{array}{l}
0 \\
a_1 \\
a_3 \\
\vdots
\end{array}
\right), \label{eq:odmat}
\\
\left( 
\begin{array}{lllll}
  \gamma_2 & \alpha_2 & & &  \\
  \epsilon_4  & \gamma_4 & \alpha_4 & &  \\
             & \epsilon_6 & \gamma_6 & \alpha_6  & \\
  & & \ddots & \ddots & \ddots 
\end{array}
\right)
\left(
\begin{array}{l}
a_1 \\
a_3 \\
a_5 \\
\vdots
\end{array}
\right)
= -
\left(
\begin{array}{lllll}
\delta & & & & \\
  & \delta & & & \\
  & & \delta & & \\
 & & & \ddots &
\end{array}
\right)
\left(
\begin{array}{l}
a_0 \\
a_2 \\
a_4 \\
\vdots
\end{array}
\right), \label{eq:evmat}
\end{eqnarray}
where  the term $\delta_n$ is constant therefore we write it
as $\delta$ here, which makes us able to accomplish the below procedure 
explicitly.
If we substitute the right hand side of Eq.(\ref{eq:evmat})
into the left hand side of Eq.(\ref{eq:odmat})
like this,
\begin{eqnarray}
\left( 
\begin{array}{lllll}
  \gamma_1 & \alpha_1 & & &  \\
  \epsilon_3  & \gamma_3 & \alpha_3 & &  \\
             & \epsilon_5 & \gamma_5 & \alpha_5  \\
   & & \ddots & \ddots & \ddots
\end{array}
\right)
\left( 
\begin{array}{lllll}
  \gamma_2 & \alpha_2 & & &  \\
  \epsilon_4  & \gamma_4 & \alpha_4 & &  \\
             & \epsilon_6 & \gamma_6 & \alpha_6  \\
 & & \ddots & \ddots &  \ddots
\end{array}
\right)
\left(
\begin{array}{l}
a_1 \\
a_3 \\
a_5 \\
\vdots
\end{array}
\right) \nonumber \\ 
= 
\left(
\begin{array}{lllll}
0 & & & & \\
\delta^2  & 0 & & & \\
  & \delta^2 & 0 & &  \\
 & & \ddots & \ddots &
\end{array}
\right)
\left(
\begin{array}{l}
a_1 \\
a_3 \\
a_5 \\
\vdots
\end{array}
\right),
\end{eqnarray}
the even sequence is eliminated and then
we can obtain a recurrence relation including 
only $b_n$ which was presented in Eq.(\ref{eq:recodd})
and the recurrence coefficients are explicitly given by
\begin{eqnarray}
\hat{\alpha}_n &=& -3n + n^2 + 8n^3 + 4n^4, \\
\hat{\beta}_n &=& 
4n( -n - 2An - 4{n^3} + nq_s + 9\rho - 24{n^2}\rho -  
    32n{\rho^2} ), \\
\hat{\gamma}_n &=& 
8 + 12A + 4{l^2(l+1)^2} + 24{n^4} - 6q_s - 4Aq_s + {q_s^2} \nonumber \\
&&  - 70\rho - 
  48A\rho + 24q_s\rho + 308{\rho^2} + 128A{\rho^2} - 64q_s{\rho^2} 
         \nonumber \\ 
&& - 
  768{\rho^3} + 1024{\rho^4} + {n^3}( -48 + 224\rho ) \nonumber \\
&& + 
  {n^2}( 46 + 16A - 8q_s - 336\rho + 864{\rho^2} )  \nonumber \\
&& + 
  2n( -11 - 8A + 4q_s + 126\rho + 48A\rho - 24q_s\rho - 
     432{\rho^2} + 768{\rho^3} ), \\
\hat{\delta}_n &=&  
8 + 4A - 16{n^4} - {{( 4 - q_s ) }^2} - 2q_s + 
  32{n^3}( 2 - 5\rho )  + 68\rho \nonumber \\
&& + 32A\rho - 16q_s\rho - 
  440{\rho^2} - 32A{\rho^2} + 16q_s{\rho^2} + 896{\rho^3} \nonumber \\
&& - 512{\rho^4} + 
  4{n^2}( -21 - 2A + q_s + 120\rho - 144{\rho^2} ) \nonumber \\
&&  + 
  4n( 10 + 4A - 2q_s - 97\rho - 8A\rho + 4q_s\rho + 288{\rho^2} - 
     224{\rho^3} ), \\
\hat{\epsilon}_n &=& 
-15n + 41{n^2} - 24{n^3} + 4{n^4} - 30\rho + 164n\rho - 
  144{n^2}\rho + 32{n^3}\rho \nonumber \\ 
 && + 164{\rho^2} - 288n{\rho^2} + 
  96{n^2}{\rho^2} - 192{\rho^3} + 128n{\rho^3} + 64{\rho^4}.
\end{eqnarray}

Similarly we get a recurrence relation for $c_n$ and
the recurrence coefficients are given by
\begin{eqnarray}
\bar{\alpha}_n &=& 1-5n^2+4n^4, \\
\bar{\beta}_n &=& 
  ( -1 + 2n ) ( 2 + 2A - 8n - 4An + 12{n^2} - 
     8{n^3}  \nonumber \\ &&
  - q_s + 2nq_s + 6\rho + 48n\rho - 48{n^2}\rho + 32{\rho^2} - 
     64n{\rho^2} ), \\
\bar{\gamma}_2 &=&
  41 + 24A + 4{A^2} - 12q_s - 4Aq_s + {q_s^2} + 306\rho  + 
   96A\rho  - 48q_s\rho  \nonumber \\
&& + 948{{\rho }^2} + 128A{{\rho }^2} - 
   64q_s{{\rho }^2} + 1536{{\rho }^3} + 1024{{\rho }^4}, \\
\bar{\gamma}_n &=& 
  38 + 24A + 4{A^2} + 24{n^4} - 12q_s - 4Aq_s \nonumber \\
&& + {q_s^2} - 308\rho - 
   96A\rho + 48q_s\rho + 956{\rho^2} + 128A{\rho^2} \nonumber \\
&& - 64q_s{\rho^2} - 
   1536{\rho^3} + 1024{\rho^4} + 32{n^3}( -3 + 7\rho ) \nonumber  \\
&& + 
   2{n^2}( 77 + 8A - 4q_s - 336\rho + 432{\rho^2} ) \nonumber \\
&&  + 
   4n( -29 - 8A + 4q_s + 189\rho + 24A\rho - 12q_s\rho - 
      432{\rho^2} + 384{\rho^3} ), \\
\bar{\delta}_2 &=&   -8 + 4A + 6q_s - {q_s^2} + 12\rho 
   - 12A\rho  + 6q_s\rho  - 
   76{{\rho }^2} \nonumber \\
&& - 16A{{\rho }^2} + 8 q_s{{\rho }^2} - 
   336{{\rho }^3} - 256{{\rho }^4},\\
\bar{\delta}_n &=& 
  -42 - 6A - 16{n^4} - {{\left( 4 - q_s \right) }^2} + 3q_s + 
   32{n^3}\left( 3 - 5\rho \right)  \nonumber \\
&& + 402\rho + 48A\rho - 24q_s\rho - 
   1160{\rho^2} - 32A{\rho^2} 
+ 16q_s{\rho^2} \nonumber \\
 && + 1344{\rho^3} - 512{\rho^4} 
 +  4{n^2}( -51 - 2A + q_s + 180\rho - 144{\rho^2} ) \nonumber  \\
&& + 
   4n( 45 + 6A - 3q_s - 247\rho - 8A\rho + 4q_s\rho + 432{\rho^2} - 
      224{\rho^3} ) ,
\\
\bar{\epsilon}_n &=& 
  21 - 76n + 83{n^2} - 32{n^3} + 4{n^4} - 152\rho + 332n\rho - 
   192{n^2}\rho \nonumber \\ 
&&+ 32{n^3}\rho + 332{\rho^2} - 384n{\rho^2} + 
   96{n^2}{\rho^2} - 256{\rho^3} + 128n{\rho^3} + 64{\rho^4}.
\end{eqnarray}

\section*{Appendix B}

We give the Gaussian elimination steps to transform the five-term
recurrence relations into three-term recurrence relations.
For $b_n$ the step defined by
\begin{eqnarray}
 \hat{\alpha}_n' & = &  \hat{\alpha}_n ,\\
 \hat{\beta}_n' & =& \hat{\beta}_n ,\\
 \hat{\gamma}_n' &=& \hat{\gamma}_n ,\\
 \hat{\delta}_n' &=& \hat{\delta}_n, \mbox{\hspace{0.5cm}for $n=1,2$},\\
 \hat{\alpha}_n' & = & \hat{\alpha}_n, \\
 \hat{\beta}_n'  & = & \hat{\beta}_n-\hat{\alpha}_{n-1}'\hat{\epsilon}_n/\hat{\delta}_{n-1}', \\
 \hat{\gamma}_n' & = & \hat{\gamma}_n- \hat{\beta}_{n-1}'\hat{\epsilon}_n/\hat{\delta}_{n-1}', \\
 \hat{\delta}_n' &=& \hat{\delta}_n- \hat{\gamma}_{n-1}'\hat{\epsilon}_n/\hat{\delta}_{n-1}', \mbox{\hspace{0.5cm}for $n \ge 3$ },
\end{eqnarray}
transforms the five-term recurrence relation into a four-term recurrence relation
and the second elimination step below,
\begin{eqnarray}
 \hat{\alpha}_1'' & = &  \hat{\alpha}_1', \\
 \hat{\beta}_1'' & =& \hat{\beta}_1' ,\\
 \hat{\gamma}_1'' &=& \hat{\gamma}_1' ,\\
 \hat{\alpha}_n''& = & \hat{\alpha}_n', \\
 \hat{\beta}_n'' & = & \hat{\beta}_n'-\hat{\alpha}_{n-1}''\hat{\delta}_n'/\hat{\gamma}_{n-1}'', \\
 \hat{\gamma}_n'' & = & \hat{\gamma}_n' - \hat{\beta}_{n-1}''\hat{\delta}_n'/\hat{\gamma}_{n-1}'',   \mbox{\hspace{0.5cm}for $n \ge 2$ },
\end{eqnarray}
brings us a final three-term recurrence relation in the following,
\begin{eqnarray}
  \hat{\alpha}_n''b_{n+1} +  \hat{\beta}_{n}''b_n +  \hat{\gamma}_n''b_{n-1} = 0,  \mbox{\hspace{0.5cm}for $n \ge 1$ }.
\end{eqnarray}
For $c_n$ the successive elimination steps in the following,
\begin{eqnarray}
 \bar{\alpha}_2' & = &  \bar{\alpha}_2, \\
 \bar{\beta}_2' & =& \bar{\beta}_2, \\
 \bar{\gamma}_2' &=& \bar{\gamma}_2, \\
 \bar{\delta}_2' &=& \bar{\delta}_2, \\ 
 \bar{\alpha}_n' & = & \bar{\alpha}_n, \\
 \bar{\beta}_n'  & = & \bar{\beta}_n-\bar{\alpha}_{n-1}'\bar{\epsilon}_n/\bar{\delta}_{n-1}', \\
 \bar{\gamma}_n' & = & \bar{\gamma}_n- \bar{\beta}_{n-1}'\bar{\epsilon}_n/\bar{\delta}_{n-1}', \\
 \bar{\delta}_n' &=& \bar{\delta}_n- \bar{\gamma}_{n-1}'\bar{\epsilon}_n/\bar{\delta}_{n-1}', \mbox{\hspace{0.5cm}for $n \ge 3$ },
\end{eqnarray}
and
\begin{eqnarray}
 \bar{\alpha}_2'' & = &  \bar{\alpha}_2', \\
 \bar{\beta}_2'' & =& \bar{\beta}_2', \\
 \bar{\gamma}_2'' &=& \bar{\gamma}_2'- \alpha_1 \bar{\delta_2}'/\gamma_1, \\
 \bar{\alpha}_n''& = & \bar{\alpha}_n', \\
 \bar{\beta}_n'' & = & \bar{\beta}_n'-\bar{\alpha}_{n-1}''\bar{\delta}_n'/\bar{\gamma}_{n-1}'', \\
 \bar{\gamma}_n'' & = & \bar{\gamma}_n' - \bar{\beta}_{n-1}''\bar{\delta}_n'/\bar{\gamma}_{n-1}'',   \mbox{\hspace{0.5cm}for $n \ge 2$ },
\end{eqnarray}
give us a three-term recurrence relation,
\begin{eqnarray}
  \bar{\alpha}_n''c_{n+1} +  \bar{\beta}_{n}''c_n +  \bar{\gamma}_n''c_{n-1} = 0, \mbox{\hspace{0.5cm}for $n \ge 2$ }. \label{eq:rec-e}
\end{eqnarray}

\section*{Appendix C}

Now we summarize here the relation between $V_1$ for multi-pole
index $l$ and $V_2$ for multi-pole index $l+1$.
If we introduce a new function $f$,
\begin{equation}
 f = \frac{r-1}{r^2},
\end{equation}
which vanishes at both boundaries, $r\rightarrow 1$ and 
$r\rightarrow \infty$,
then two potential are given by
\begin{eqnarray}
  V_1 &=& + (l+1) \frac{df}{dr_*}
-4  f^3 + (l+1)^2 f^2, \label{eq:v1} \\
  V_2 &=& - (l+1) \frac{df}{dr_*}
-4  f^3 + (l+1)^2 f^2. \label{eq:v2}
\end{eqnarray}
If these two different potentials yield the same amplitudes of
transmission and reflection, the following integrals for two potentials
should be the same;
\begin{eqnarray}
I_1 &=& \int V dr_* ,\\
I_2 &=& \int V^2 dr_* ,\\
I_3 &=& \int (2V^3 +V'^2) dr_* ,\\
I_4 &=& \int (5V^4+10VV'^2+V''^2) dr_* ,\\
I_5 &=& \int (14V^5+70V^2V'^2+14VV''^2+V'''^2) dr_*, \\
I_6 &=& ... \nonumber 
\end{eqnarray}
Due to the simple forms, Eq.(\ref{eq:v1}) and Eq.(\ref{eq:v2}),
we can see the coincidence of these integrals, $I_1 \sim I_{10}$,
for two potentials.
The explicit forms of above integrals, $I_1 \sim I_{5}$, are given by 
\begin{eqnarray}
I_1 &=& (1+6l+3l^2)/3,\\
I_2 &=& (4 + 18l + 93{l^2} + 84{l^3} + 21{l^4})/ {630},\\
I_3 &=& (23 + 120l + 476{l^2} + 1560{l^3}  \nonumber \\
    && +1820{l^4} + 858{l^5} +  143{l^6})/ {45045} \\
I_4 &=& (1648 + 8736l + 32716{l^2} + 105868{l^3} \nonumber \\
&&  + 252567{l^4} +  316540{l^5} + 203490{l^6} \nonumber \\
&& + 64600{l^7} + 8075{l^8})/  {9399380},\\
I_5 &=&  
  (25160 + 125820l + 437634{l^2} + 1336860{l^3} \nonumber \\
&& + 3430245{l^4} + 
      6563970{l^5} + 8099427{l^6} + 6085800{l^7} \nonumber \\
&& + 2687895{l^8} + 
      642390{l^9} + 64239l^{10})/ {1003917915}. 
\end{eqnarray}

\newpage
\begin{table}
\begin{center}
\begin{tabular}{|l|l|l|l|}
\hline
      & $s=2, l=2$          & $s=2, l=3$           & $s=2, l=4$  \\ \hline
n=0   & (-0.083460,0.43134) & (-0.085973,0.70430)  & (-0.087001,0.96576) \\
Leaver& (-0.083645,0.43098) &                      &                     \\
WKB   & (-0.08349,0.43013)  & (-0.08596,0.70398)   & (-0.08700,0.96563)  \\
n=1   & (-0.25498,0.40452)  & (-0.25992,0.68804)   & (-0.26212,0.95381)  \\
Leaver& (-0.257055,0.39309) &                      &                     \\
WKB   & (-0.25675,0.40076)  & (-0.26014,0.68701)   & (-0.26218,0.95339)  \\
n=2   & (-0.44137,0.35340)  & (-0.44007,0.65624)   & (-0.44064,0.93020)  \\
Leaver& (-0.442035,0.353515)&                      &                     \\
WKB   & (-0.44210,0.35136)  & (-0.43986,0.65575)   & (-0.44056,0.93004)  \\ \hline
      &  $s=1, l=1$         &  $s=1, l=2$          & $s=1, l=3$ \\ \hline 
n=0   & (-0.083460,0.43134) & (-0.085973,0.70430)  & (-0.087001,0.96576) \\
Leaver& (-0.08343,0.431415) & (-0.086205,0.704075) &                     \\
WKB   & (-0.08349,0.43013)  & (-0.08596,0.70398)   & (-0.08700,0.96563)  \\
n=1   & (-0.25498,0.40452)  & (-0.25992,0.68804)   & (-0.26212,0.95381)  \\
Leaver& (-0.259705,0.40602) & (-0.26256,0.68315)   &                     \\
WKB   & (-0.25675,0.40076)  & (-0.26014,0.68701)   & (-0.26218,0.95339)  \\
n=2   & (-0.44137,0.35340)  & (-0.44007,0.65624)   & (-0.44064,0.93020)  \\
Leaver& (-0.44260,0.35347)  & (-0.4408,0.655675)   &                     \\
WKB   & (-0.44210,0.35136)  & (-0.43986,0.65575)   & (-0.44056,0.93004) \\ \hline
     &  $s=0, l=0$         &  $s=0, l=1$           & $s=0, l=2$ \\ \hline 
n=0  & (-0.095844,0.13346) & (-0.089384,0.37764)   & (-0.088748,0.62657) \\
WKB  & (-0.10371,0.12109)  & (-0.08936,0.37570)    & (-0.08873,0.62609)  \\
n=1  & (-0.33065,0.092965) & (-0.27614,0.34818)    & (-0.26909,0.60817)  \\
WKB  & (-0.33742,0.09157)  & (-0.27828,0.34392)    & (-0.26944,0.60677)  \\
n=2  & (-0.58833,0.075081) & (-0.48643,0.29846)    & (-0.45820,0.57287)  \\ 
WKB  & (-0.57164,0.05056)  & (-0.48145,0.29661)    & (-0.45750,0.57254)  \\ \hline
\end{tabular} 
\end{center}
\caption{The computed quasinormal frequencies are listed.
Our results are very consistent with the values of WKB quasinormal
frequencies.  The relative errors of WKB complex and real
gravitational quasinormal frequencies for $l=2$ from our results
are (0.04\%, 0.03\%), (0.07\%, 0.1\%) and (0.1\%,0.5\%) for $n=0,1,2$,
respectively.
Leaver's value has rather large discrepancy of 3\% for the real
frequency of $n=1$ gravitational wave.}
\end{table}

\begin{figure}
\begin{center}
  \leavevmode
  \epsfysize=9.0cm
  \epsfbox{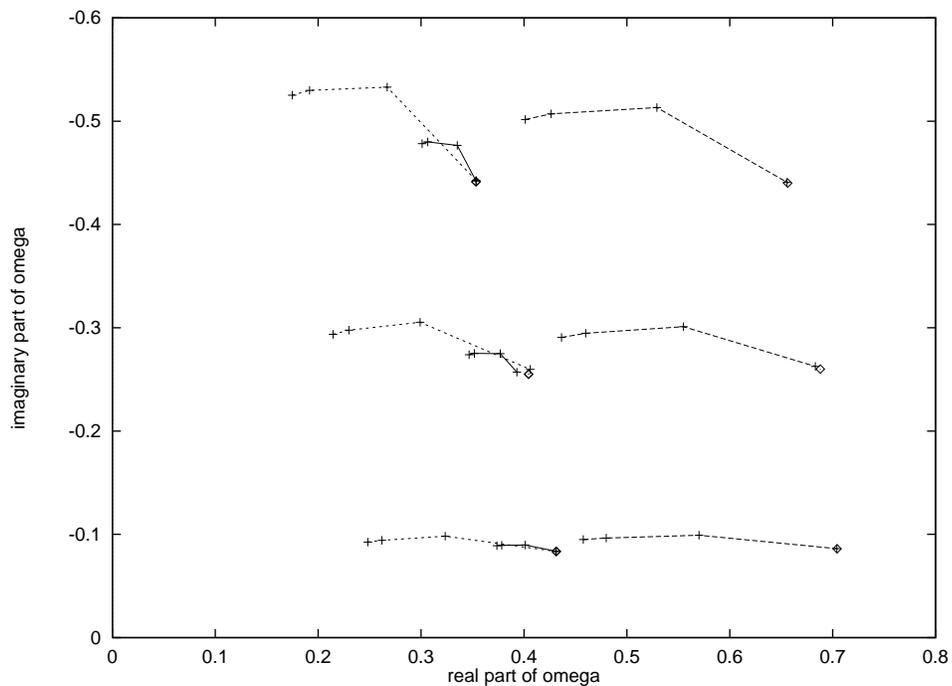}
\end{center}
\caption{
Quasinormal frequencies for  
$l=2$ gravitational wave and $l=1,2$ electromagnetic wave
are plotted in the $\omega$-plane with the results of Leaver[13].
Each solid line is a trajectory of the first three lowest 
$l=2$ gravitational mode, parameterized by the charge $Q$.
It has a tendency to coincide at the right end
with $l=1$ electromagnetic mode which is shown as
a small dashed line.
Marks from left to right correspond to $Q=0,0.4,0.8,0.9999$
quasinormal frequencies of Leaver.
The frequencies of extremal black holes we computed are plotted as diamonds.
Dashed lines are $l=2$ electromagnetic modes.
}
\end{figure}

\begin{figure}
\begin{center}
  \leavevmode
  \epsfysize=9.0cm
  \epsfbox{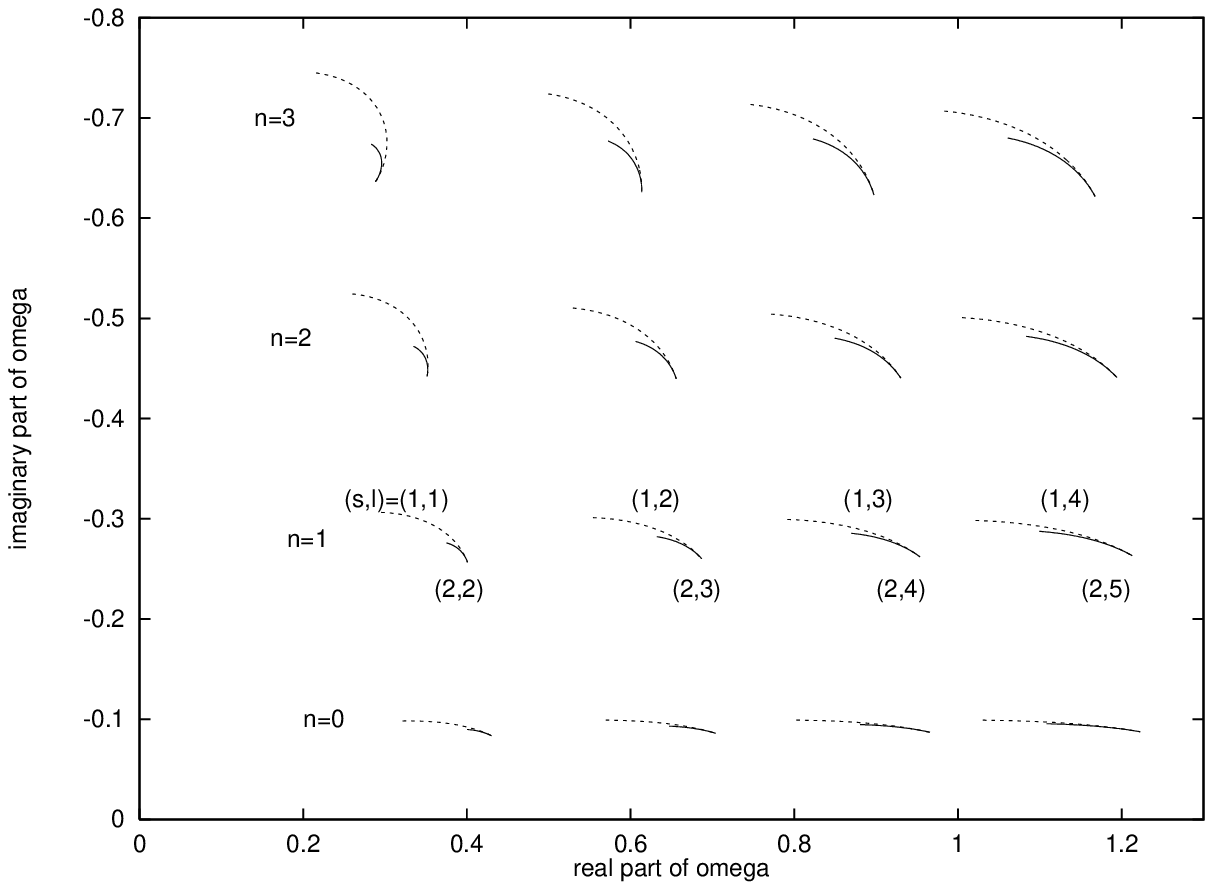}
\end{center}
\caption{
Solid lines and dashed lines are trajectories of the third order WKB
quasinormal frequencies of the gravitational and electromagnetic wave,
respectively.
Each left endpoint of lines corresponds to the quasinormal frequency
of a charged black hole of $Q=0.8$, and each right endpoint corresponds to
the frequency in the limit of maximal charge.
A trajectory of the gravitational quasinormal frequencies
depicted in a solid line
meets at the right end with a corresponding dashed line,
which is a trajectory of the electromagnetic quasinormal
frequencies belonging to lower multi-pole index by one.
}
\end{figure}


\begin{thebibliography}{10}

\bibitem{REG57}
T.~Regge and J.~Wheeler,
\newblock {\em Phys. Rev.} {\bf 108} (1957) 1063.

\bibitem{ZER71}
F.~Zerilli,
\newblock {\em Phys. Rev. D} {\bf 2} (1971) 2141.

\bibitem{ZER74}
F.~Zerilli,
\newblock {\em Phys. Rev. D} {\bf 9} (1974) 860.

\bibitem{MON}
V.~Moncrief,
\newblock {\em Phys. Rev. D} {\bf 9} (1974) 2707; {\it ibid.}
{\bf 10} (1974) 1057; {\it ibid.} {\bf 12} (1975) 1526.

\bibitem{TEU72}
S.~Teukolsky,
\newblock {\em Phys. Rev. Lett.} {\bf 29} (1972) 1114.

\bibitem{CHA75a}
S.~Chandrasekhar and S.~Detweiler,
\newblock {\em Proc. Roy. Soc. London} {\bf A344} (1975) 441.

\bibitem{LEA85}
E.~Leaver,
\newblock {\em Proc. Roy. Soc. London} {\bf A402} (1985) 285.

\bibitem{GAU67}
W.~Gautschi,
\newblock {\em SIAM Rev} {\bf 9} (1967) 24.

\bibitem{FER84}
V.~Ferrari and B.~Mashhoon,
\newblock {\em Phys. Rev. D} {\bf 30} (1984) 295.

\bibitem{BOL84}
H.~Blome and B.~Mashhoon,
\newblock {\em Phys. Lett.} {\bf 100A} (1984) 231.

\bibitem{SCH85}
B.~Schutz and C.~Will,
\newblock {\em Astrophys. J.} {\bf 291} (1985) L33.

\bibitem{IYE87}
S.~Iyer and C.~Will,
\newblock {\em Phys. Rev. D} {\bf 35} (1987) 3621;
S.~Iyer,
\newblock {\it ibid.} {\bf 35} (1987) 3632;
E.~Seidel and S.~Iyer,
\newblock {\it ibid.} {\bf 41} (1990) 374.

\bibitem{LEA90}
E.~Leaver,
\newblock {\em Phys. Rev. D} {\bf 41} (1990) 2986.

\bibitem{KOK88}
K.~Kokkotas and B.~Schutz,
\newblock {\em Phys. Rev. D} {\bf 37} (1988) 3378.

\bibitem{CHA83}
S.~Chandrasekhar,
\newblock {\em The Mathematical Theory of Black Holes},
\newblock (Clarendon, Oxford, 1983)

\bibitem{TEU74}
S.~Teukolsky and W.~Press,
\newblock {\em Astrophys. J.} {\bf 193} (1974) 443.

\bibitem{DET80}
S.~Detweiler,
\newblock {\em Astrophys. J.} {\bf 239} (1980) 292;
\newblock {\em Phys. Rev. Lett.} {\bf 52} (1983) 67.

\bibitem{SAS90}
M.~Sasaki and T.~Nakamura,
\newblock {\em Gen. Rel. Grav.} {\bf 22} (1990) 1351.

\end{thebibliography}
\end{document}